\documentclass[journal=jpclcd,manuscript=letter]{achemso}
\usepackage{chemformula}
\usepackage[T1]{fontenc} 
\usepackage{xcolor,color}
\usepackage{verbatim}
\usepackage{float}
\usepackage{mathrsfs}
\usepackage{amsmath}
\usepackage{amssymb}
\usepackage{graphics}
\usepackage{graphicx}
\usepackage{physics}

\def\di{{\rm d}}
\def\eu{{\rm e}}
\def\iu{{\rm i}}

\def\expv#1{\left\langle #1 \right\rangle}
\def\proj#1#2{\left| #1 \left\rangle \right\langle #2\right|}
\def\bok#1#2#3{\left\langle#1\left|#2\right|#3 \right\rangle}

\def\ket#1{|\,#1\, \rangle}

\usepackage{microtype}
\usepackage[hyperindex=true,pdftitle={Quantum to Classical Cavity Chemistry Electrodynamics},
colorlinks=true,pagebackref=false,citecolor=blue,plainpages=false,
pdfpagelabels, breaklinks=true,linkcolor=black,urlcolor=blue]{hyperref}

\def\tnP{\textnormal{P}} 
\def\tnM{\textnormal{M}} 
\def\tnS{\textnormal{m}} 
\def\tnR{\textnormal{ph}} 
\def\tnSR{\textnormal{ph-m}} 
\def\tnq{\textnormal{q}} 
\def\tnp{\textnormal{p}} %
\def\suPW{\textnormal{W}} 
\def\suW{\textnormal{W}}  
\def\suI{\textnormal{I}} 

\usepackage[symbol]{footmisc}

\def\tr{\mathrm{tr}}

\newcommand{\GFTyMA}{
Grupo de F{\'i}sica Te{\'o}rica y Matem{\'a}tica Aplicada, Instituto de
F{\'i}sica, Facultad de Ciencias Exactas y Naturales, 
Universidad de Antioquia; Calle 70 No.~52-21, Medell\'in, Colombia.}

\newcommand{\FICOMACO}{
Grupo de F\'{\i}sica Computacional en Materia Condensada, 
Escuela de F\'{\i}sica, Facultad de Ciencias, 
Universidad Industrial de Santander UIS; Cra 27 Calle 9 
Ciudad Universitaria, Bucaramanga, Colombia.}

\newcommand{\GFAM}{
Grupo de F\'isica At\'omica y Molecular, Instituto de F\'{\i}sica,  
Facultad de Ciencias Exactas y Naturales, Universidad de Antioquia UdeA;
Calle 70 No. 52-21, Medell\'in, Colombia.}

\author{Leonardo F. Calder\'on}
\email{leonardo.calderon@utoronto.ca}
\affiliation{\GFTyMA}
\alsoaffiliation{\FICOMACO}
\author{Humberto Trivi\~no}
\affiliation{\GFTyMA}
\author{Leonardo A. Pach\'on}
\email{leonardo.pachon@udea.edu.co}
\affiliation{\GFTyMA}
\alsoaffiliation{\GFAM}
\title[]
{Quantum to Classical Cavity Chemistry Electrodynamics}
\date{\today}

\begin{document}

\begin{tocentry}
\includegraphics[scale=0.5]{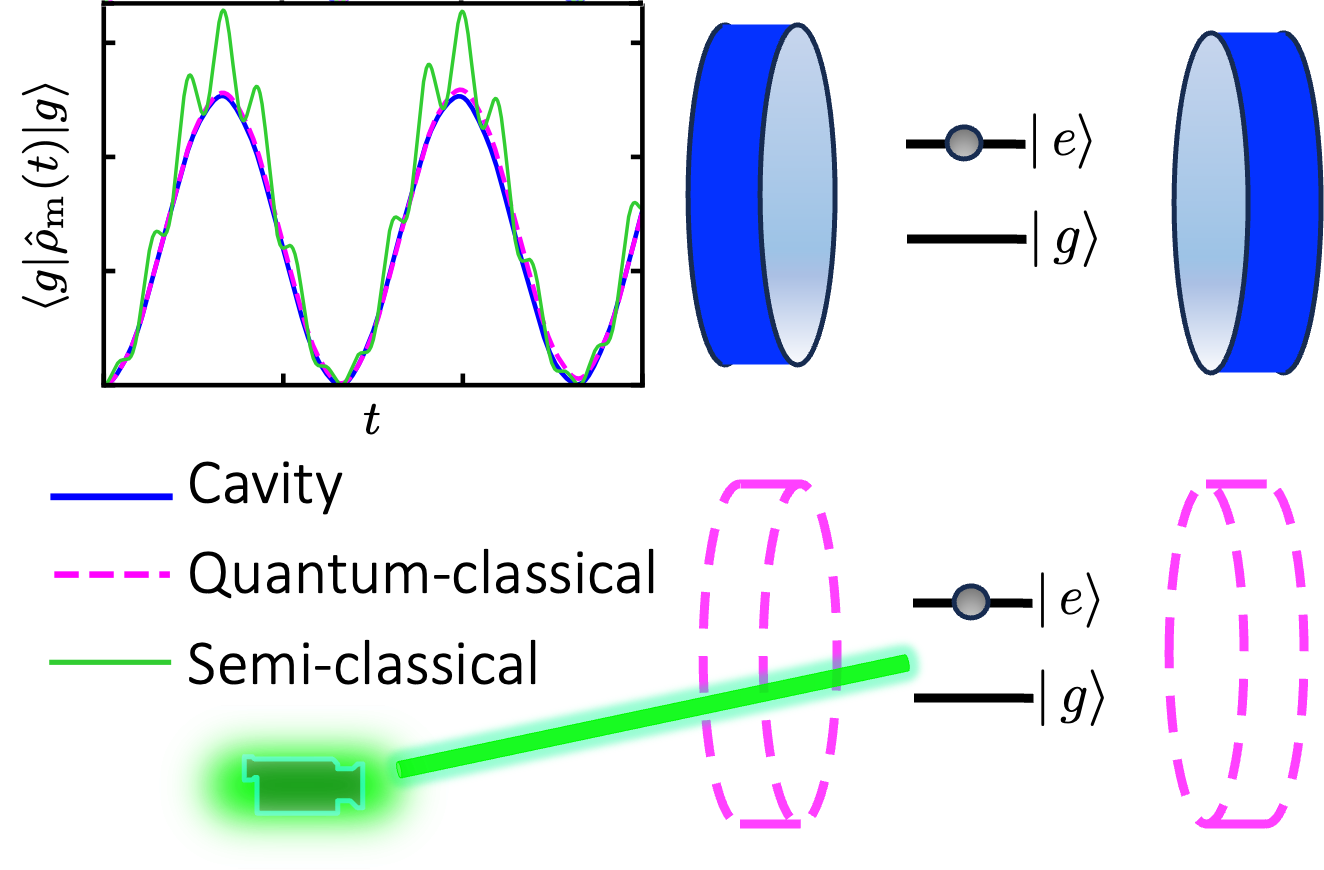}
\end{tocentry}

\begin{abstract}
Polaritonic chemistry has ushered in new avenues for controlling molecular 
dynamics. 
However, two key questions remain: (i) Can classical light sources elicit the 
same effects as certain quantum light sources on molecular systems? (ii) Can 
semiclassical treatments of light-matter interaction capture nontrivial quantum 
effects observed in molecular dynamics? 
This work presents a quantum-classical approach addressing issues of realizing 
cavity chemistry effects without actual cavities. 
It also highlights the limitations of the standard semiclassical light-matter 
interaction. 
It is demonstrated that classical light sources can mimic quantum effects up 
to the second order of light-matter interaction, provided that the mean-field 
contribution, symmetrized two-time correlation function, and the linear 
response function are the same in both situations. 
Numerical simulations show that the quantum-classical method aligns more 
closely with exact quantum molecular-only dynamics for quantum light states 
such as Fock states, superpositions of Fock states, and vacuum squeezed 
states than the conventional semiclassical approach.
\end{abstract}

\textbf{\textit{Introduction}}.\textemdash
Quantum cavity chemistry explores the potential to transform the chemical
landscape, paving the way for novel photophysical phenomena 
\cite{Ebb16,RM&18,RT&18,FRN18,DR&18,HS18,FGG18,KM&19,HW&19,HO20,MT&23,CP&23}. 
This transformation is achieved by coupling molecular systems to the 
confined photonic degrees of freedom within microcavities\cite{KB&07}. 
As a result, it offers a promising avenue for manipulating 
room-temperature photophysical processes in organic molecules. 
Instances of these novel phenomena have been examined in the context of 
singlet fission processes \cite{MD&18}, triplet harvesting \cite{ME&19}, 
energy transfer \cite{ZC&16,ZC&17,SF&18,DM&18,SF&19}, remote regulation 
of chemical reactions \cite{DRY19}, and nonadiabatic effects \cite{GGF15,KBM16,GM20}.

In classical control theory\cite{SB03}, properties of light, such as 
intensity and phase, are employed to alter the inherent dynamics of 
molecular reactions using classical laser-control schemes. 
In contrast, quantum cavity chemistry leverages the quantum nature of 
light as a vital tool for reshaping the chemical behavior of molecular 
reactions. 
However, despite the nonadiabatic effects observed in the adiabatic 
ground state populations, evidence from the \textnormal{NaI} molecule 
\cite{KBM16,CH&17} suggests minimal discernible differences between 
dynamics driven by classical laser radiation and those induced by cavity 
interactions. 
This observation has piqued interest in exploring how classical 
radiation fields might replicate the molecular dynamics typically 
attributed to quantum light states\cite{CV&19}.

Traditionally, in the study of light-matter interactions \cite{CP20,CCB23}, 
semiclassical methods have involved a full quantum treatment of the molecular 
degrees of freedom, while adopting various classical dynamics schemes for the 
photonic degrees of freedom \cite{HS&19,HS&19a,CL&19,LC&20,RH&22,HW&22,LTH22,HKT23a,HKT23b}. 
However, recent advances employing mixed quantum-classical dynamics 
techniques \cite{Ger82,GKK09} to solve the Ehrenfest mean-field dynamics 
have made strides. These techniques have demonstrated an ability to 
account, at least partially, for phenomena such as spontaneous emission, 
interference, strong coupling, and correlated light-matter dynamics 
\cite{HS&19,HS&19a,CL&19,LC&20,RH&22,HW&22,LTH22,HKT23a,HKT23b}.

Utilizing a mixed representation of molecule operators and radiation 
phase-space distributions\cite{KC99,Kap15,TK01}—and by tracing out the radiation 
degrees of freedom—an effective master equation exclusive to the molecular 
degrees of freedom was derived. 
To achieve the closeness condition for this master equation, contributions 
were considered up to the second order in light-matter interaction. 
This closed master equation facilitates a direct juxtaposition of the full 
quantum dynamics, the conventional semiclassical light-matter interaction, 
and an alternative semiclassical description that is non-local in time. 
It should be noted that the primary focus of this paper is not on 
demonstrating the application of trajectory methods \cite{HS&19a,LC&20,RH&22}. 
Instead, the emphasis is on elucidating the impact of photonic degrees of 
freedom on molecular-only dynamics, framed in the context of radiation 
field statistics.

In a nutshell, up to the second order in light-matter interaction, quantum
effects can be mimicked by classical light sources if the mean-field 
contribution, the symmetrized two-time correlation function, and the 
linear response function (defined through the antisymmetrized two-time 
correlation function) are consistent across both scenarios. 
This condition does not pertain to the classical limit of the quantum 
light source. 
Instead, it highlights the potential to design classical light sources 
that mimic the radiation’s effect on the molecular degrees of freedom up 
to the second order.

In this context, the impact of quantum fluctuations from three distinct 
quantum light field states—a Fock state, a Fock state 
superposition, and a squeezed vacuum state—is analyzed using system 
models. 
These states possess a non-trivial quantum character\cite{MW95,GK05,Aga12} 
and facilitate the exploration of the influence of both vanishing and 
non-vanishing mean-field contributions and the 
symmetrized/antisymmetrized two-time correlation functions on 
molecular-only dynamics.

This letter is structured as follows. 
First, a description of a molecular system coupled to a 
quantized light field is provided, accompanied by the formulation of a 
second-order quantum master equation that treats the light degrees of 
freedom quantum mechanically. 
Then, a quantum-classical projection operator formalism is introduced, from 
which a quantum-classical second-order master equation is derived for the 
molecular system, considering the light degrees of freedom classically.
Lastly, numerical results that juxtapose the quantum-classical, standard 
semiclassical, and precise quantum dynamics of Hamiltonians employed in 
theoretical molecular polaritonics are presented, 
considering various light field states with nonclassical characteristics.

\textbf{\textit{Quantum cavity molecular-only dynamics}}.\textemdash
Consider a molecular system that interacts with a quantized light field. 
Assuming the dipole approximation is valid, the interaction is framed in 
the length gauge \cite{FR&17,RT&18}. 
In this configuration (Pauli-Fierz model), the correlated 
electron-nuclear-photon Hamiltonian, which 
comprises $n_{\mathrm{el}}$ electrons, 
$n_{\mathrm{nul}}$ nuclei, and $n_{\mathrm{ph}}$ photon modes, can be 
expressed as a sum of the electro-nuclear (or bare molecular) Hamiltonian, 
denoted as $\hat{H}_{\mathrm{bm}}$, and the photon Hamiltonian, represented 
as $\hat{H}_{\mathrm{ph}}$. 
The bare molecular Hamiltonian, $\hat{H}_{\mathrm{bm}}$, includes five 
components: the kinetic energy of the electrons $\hat{T}_{\mathrm{el}}$, 
the kinetic energy of the nucleons $\hat{T}_{\mathrm{nuc}}$, the 
interaction between electrons $ \hat{V}_{\mathrm{el-el}}$, between 
nucleons $\hat{V}_{\mathrm{nuc-nuc}}$ and the electron-nucleons 
interaction $\hat{V}_{\mathrm{el-nuc}}$, it reads
$
\hat{H}_{\mathrm{bm}} 
= \hat{T}_{\mathrm{el}} +  \hat{T}_{\mathrm{nuc}} +
\hat{V}_{\mathrm{el-el}}  +  \hat{V}_{\mathrm{nuc-nuc}} + 
\hat{V}_{\mathrm{el-nuc}}$. 
The photon Hamiltonian incorporates the coupling to the molecular degrees
of freedom and reads 
\begin{equation}
\begin{split}
\hat{H}_{\mathrm{ph}} &= \frac{1}{2}\sum_{\alpha=1}^{n_\mathrm{ph}}
\Big[\hat{\tnp}_{\alpha}^2 + \omega_{\alpha}^2 
\big(\hat{\tnq}_{\alpha} +e\;\boldsymbol{\lambda}_{\alpha} 
\cdot \hat{\mathbf{R}}/\omega_{\alpha} \big)^2\Big]
\\
&= \hat{H}_{\mathrm{bph}} + \hat{H}_{\tnSR} + \hat{H}_{\mathrm{rm}},
\end{split}
\end{equation} 
with $\hat{\mathbf{R}} 
= \sum_{I=1}^{n_{\mathrm{nuc}}} Z_I \hat{\mathbf{X}}_I -
\sum_{i=1}^{n_{\mathrm{el}}} \hat{\mathbf{x}}_i$ 
being the total dipole operator including both sets of electronic 
$\{\hat{\mathbf{x}}_i\}$ and nuclear $\{\hat{\mathbf{X}}_I\}$ coordinates
and $Z_I$ stand for the nuclear charges. 
$\hat{H}_{\mathrm{bph}}=\frac{1}{2}\sum_{\alpha=1}^{n_\mathrm{ph}}
(\hat{\tnp}_{\alpha}^2 + \omega_{\alpha}^2\hat{\tnq}_{\alpha}^2)$ 
denotes the bare photon Hamiltonian, 
$\hat{H}_{\tnSR} = 
\sum_{\alpha=1}^{n_\mathrm{ph}} \omega_{\alpha}\hat{\tnq}_{\alpha}
\boldsymbol{\lambda}_{\alpha}\cdot e \hat{\mathbf{R}}$ 
does so for the interaction between photons and molecular degrees of
freedom, with $\boldsymbol{\lambda}$ as the dipole coupling strength. 
Finally, 
$\hat{H}_{\mathrm{rm}} = 
\frac{1}{2}\sum_{\alpha=1}^{n_\mathrm{ph}} 
\big(\boldsymbol{\lambda}_{\alpha} \cdot e \hat{\mathbf{R}} \big)^2$
accounts for the renormalization of the bare molecular Hamiltonian due to
the coupling to light and corresponds to the dipole self-energy. 

Another more conventional formulation for studying molecule-cavity physics 
corresponds to the multipolar macroscopic model of QED under the 
Power-Zineau-Woolley (PZW) unitary transformation from the minimal 
coupling \cite{PZ59,AW70,Woo71}. 
Macroscopic QED allows for quantizing the electromagnetic field in any 
geometry, considering realistic cavities with lossy materials 
\cite{SC08,FFG20,ST&23}. 
Light-matter interactions are addressed directly through the electric 
$\mathbf{E}$ and magnetic $\textbf{B}$ fields, and the ﬁeld-emitter 
interactions are incorporated by multipole moments \cite{SC08,FFG20}. 
In multipolar QED, the dipole self-energy contribution is contained in 
the molecular Hamiltonian and does not directly depend on the vacuum 
field \cite{FFG20,HH19,RW&18,MT&23}. 
In addition, for realistic cavities where the interaction is driven mainly by 
longitudinal fields (e.g., plasmonic nanocavities), the dipole self-energy 
does not appear \cite{FFG20,FGF22}. 

To provide a general insight into the quantum aspects of light in
molecular systems, consider the molecular system-light field Hamiltonian 
\begin{equation}
\label{equ:totalH}
\hat{H} = \hat{H}_{\mathrm{m}} +  \hat{H}_{\tnSR} + \hat{H}_{\mathrm{bph}} ,
\end{equation} 
with the molecular Hamiltonian 
$\hat{H}_{\mathrm{m}}=\hat{H}_{\mathrm{bm}} + \hat{H}_{\mathrm{rm}}$. 
The dynamics of the state $\hat{\rho}$ of the entire system described by
$\hat{H}$ follows from the von Neumann equation, 
$\frac{\mathrm{d}}{\mathrm{d} t}\hat{\rho} = -\frac{\mathrm{i}}{\hbar}
\big[\hat{H},\hat{\rho}\big]$. 
Because interest here is in the influence that coupling to photons induces
on the molecular systems, the partial trace over the states of the photons
is applied to $\hat{\rho}$, so that the reduced state of the molecular
systems follows from 
$\hat{\rho}_{\mathrm{m}} = \mathrm{tr}_{\mathrm{ph}} \hat{\rho}$. 
To focus exclusively on the dynamics induced by the light-matter coupling,
it is convenient to calculate the equation of motion of 
$\hat{\rho}_{\mathrm{m}}$ in the interaction picture by introducing the
unitary transformation 
$\hat{\rho}^{\mathrm{(\suI)}}(t) = \hat{U}^\dagger(t-t_0) 
\hat{\rho}(t_0) \hat{U}(t-t_0) $ 
with $\hat{U}(t-t_0) = \exp \big[\frac{\mathrm{i}}{\hbar}
\hat{H}_{\mathrm{m}}(t-t_0) \big]
\exp \left[\frac{\mathrm{i}}{\hbar} \hat{H}_{\mathrm{bph}}(t-t_0) \right]$.
Therefore, 
\begin{equation}
\label{eq:von_Neumann_interaction}
\begin{split}
\frac{\mathrm{d}}{\mathrm{d} t}\hat{\rho}^{\mathrm{(\suI)}}_{\mathrm{m}}(t)&=
-\frac{\mathrm{i}}{\hbar} \mathrm{tr}_{\mathrm{ph}} 
\left[\hat{H}^{\mathrm{(\suI)}}_{\tnSR}(t),\hat{\rho}^{\mathrm{(\suI)}}(t)\right].
\end{split}
\end{equation} 
By continuing the discussion in the interaction picture, the sheer
complexity of the intrinsic molecular dynamics is left out, and the
light-matter interaction becomes the focus of the description below.

\textbf{\textit{Quantum-classical cavity molecular-only dynamics}}.\textemdash 
Once the light fields have been traced out, the core proposition of this 
work emerges: if the reduced equation of motion for 
$\hat{\rho}_{\mathrm{m}}$ remains consistent under both quantum and 
classical light, then it becomes feasible to \emph{effectively} simulate 
the molecular dynamics triggered by quantum light fields using classical 
counterparts. 
To further explore the capability of classical fields to replicate the 
dynamics stimulated by quantum fields, the radiation degrees of freedom 
will be translated into phase-space distributions in the subsequent 
sections. 

To achieve this, an initial step involves executing a partial Wigner-Weyl 
transform\cite{Wey27,Wig32} on the light field's Hilbert space. 
This process provides a phase-space depiction of the light field's degrees 
of freedom. 
The partial Wigner-Weyl representation for the state $\hat{\rho}$ is given as follows: 
$
\label{eq:def_wigner_function}
\hat{\rho}^{\suPW}(\mathbf{q},\mathbf{p}) = 
\frac{1}{(2\pi\hbar)^{n_{\mathrm{ph}}}}\int\di 
\mathbf{u}
\langle{\mathbf{q} + \frac{\mathbf{u}}{2}}|\hat{\rho}|\mathbf{q}-\frac{
\mathbf{u}}{2}\rangle \eu^{-\frac{\iu}{\hbar}\mathbf{p}\cdot \mathbf{u}},
$ 
where the integration is performed over the entire phase-space of light
field degrees of 
freedom $(\mathbf{q},\mathbf{p})=(\{\tnq_{\alpha}\},\{\tnp_{\alpha}\})$. 
The state $\hat{\rho}^{\suPW}$ retains its operator nature since the
molecular degrees of freedom still represent quantum operators in the
molecule Hilbert space $\mathcal{H}_\mathrm{m}$. 
After performing the Wigner-Weyl transform, the molecule-light field
dynamics is governed by the von-Neumann--Moyal equation \cite{Moy49,BS13} 
\begin{equation}
\label{equ:MoyalBracket}
\frac{\partial\hat{\rho}^{\suPW}}{\partial t} =  
\left\{\hat{H}^{\suPW},\hat{\rho}^{\suPW}\right\}_{\tnM}
= -\frac{\iu}{\hbar} ( 
\hat{H}^{\suPW}\eu^{\frac{\iu\hbar}{2}\Lambda}\hat{\rho}^{\suPW}
- \hat{\rho}^{\suPW}\eu^{\frac{\iu\hbar}{2}\Lambda}\hat{H}^{\suPW} ),
\end{equation}
where $\{ \cdot, \cdot \}_{\mathrm{M}}$ symbolizes the Moyal bracket, with 
$\eu^{\frac{\iu\hbar}{2}\Lambda}=\lim\limits_{N\to\infty} 
\sum_{l=0}^{N}\frac{1}{l!}
(\frac{\iu\hbar}{2}\Lambda)^l$, and 
$\Lambda = 
\sum_{\alpha=1}^{n_{\mathrm{ph}}}\frac{\overleftarrow{\partial}}
{\partial \tnq_{\alpha}}
\frac{\overrightarrow{\partial}}{\partial
\tnp_{\alpha}}
- \frac{\overleftarrow{\partial}}
{\partial \tnp_{\alpha}}\frac{\overrightarrow{\partial}}{\partial
\tnq_{\alpha}}$. 
The arrows indicate the direction in which a derivative acts. 
In this representation there is no approximation invoked.

The classical limit for the light degrees of freedom of 
Eq.~(\ref{equ:MoyalBracket}) corresponds to the von-Neumann--Poisson
equation that is reached by disregarding $\mathcal{O}(\hbar^2)$ terms in the
expansion of the phase-space operator $\eu^{\frac{\iu\hbar}{2}\Lambda}$. 
Thus, the quantum-classical dynamics \cite{KC99,Kap15} of the state
$\hat{\rho}^{\suPW}$ reads 
\begin{equation}
\label{eq:qcle}
\frac{\partial \hat{\rho}^{\suPW}}{\partial t} = 
-\iu\mathcal{L}^{\suPW}\hat{\rho}^{\suPW} 
= -\frac{\iu}{\hbar}\left[ \hat{H}^{\suPW},\hat{\rho}^{\suPW}\right] 
   + \frac{1}{2}\left( 
\left\{\hat{H}^{\suPW},\hat{\rho}^{\suPW}\right\}_{\tnP} - 
\left\{\hat{\rho}^{\suPW},\hat{H}^{\suPW}\right\}_{\tnP} 
\right) ,
\end{equation} 
where $\{ \cdot, \cdot \}_{\mathrm{P}}$ denotes the classical Poisson 
bracket. 
It is essential to point out that the dynamics generated by the
quantum-classical Liouvillian superoperator $\mathcal{L}_{\suPW}$ coincides
with the Moyal bracket's quantum dynamics as the photon Hamiltonian is
quadratic in its quadratures. 
In this partial Wigner representation, the Hamiltonian of the 
Eq.~(\ref{equ:totalH}) reads 
\begin{equation}
\label{eq:total_Hamiltonian_pW}
\hat{H}^{\suPW} = \hat{H}_{\tnS} + \hat{H}_{\tnSR}^{\suPW} + 
H_{\tnR}^{\suW}.
\end{equation} 
The interaction between the light field and the molecular degrees of
freedom
is defined by the Hamiltonian of the general form 
$\hat{H}_{\tnSR}^{\suPW} = \sum_{u} \hat{K}_u\Phi_{u}^{\suW}$, where
$\hat{K}_u$ represents an operator in the Hilbert space of the molecular
system, and $\Phi_{u}^{\suW}$ the phase-space representation of operator
$\hat{\Phi}_{u}$ of the light field. 
Note that the photon Hamiltonian $H_{\tnR}^{\suW}$ is not an operator
anymore.

The partial trace over the light degrees of freedom is obtained as 
$\hat{\rho}_{\tnS} = \tr_{\tnR} \hat{\rho}^{\suPW}
= \int \di \mathbf{q} \di \mathbf{p} \hat{\rho}^{\suPW}$. 
After replacing the Hamiltonian of the Eq.~(\ref{eq:total_Hamiltonian_pW}) 
in the Eq.~(\ref{eq:qcle}), and considering that  
$\big[ H_{\tnR}^{\suW},\hat{\rho}^{\suPW}\big] 
=
\big\{ \hat{H}_{\tnS}, \hat{\rho}^{\suPW}\big\}_{\tnP} = 0 $, 
the evolution of the reduced state of the molecular system reads 
\begin{equation}
\label{equ:PWIRdcdMstrEqtn}
\begin{split}
&\frac{\partial \hat{\rho}_{\tnS}}{\partial t} = 
- \frac{\iu}{\hbar}\left[ \hat{H}_{\tnS},\hat{\rho}_{\tnS}\right] 
- \frac{\iu}{\hbar}\sum_{u}\left[ \hat{K}_u,\tr_{\tnR}
\left(\Phi_{u}^{\suW}\hat{\rho}^{\suPW}\right) \right]
\\ &
\!+ \frac{1}{2}\!\sum_{u}\!
\left[ 
\hat{K}_u,\tr_{\tnR}
\left\{\Phi_{u}^{\suW},\hat{\rho}^{\suPW}\right\}_{\tnP}\!
\right]_{\boldsymbol{+}}
\!+\!\tr_{\tnR}\left\{ H_{\tnR}^{\suW},\hat{\rho}^{\suPW}\right\}_{\tnP}. 
\end{split}
\end{equation} 
The full influence of the quantum/classical light field into the molecular
dynamics is encoded in $\Phi_{u}^{\suW}$ provided that no approximation is
involved in Eq.~(\ref{equ:PWIRdcdMstrEqtn}). 
Therefore, deviations from the quantum description can be discussed in terms
of the quantum/classical correlations of $\Phi_{u}^{\suW}$.

\textit{Derivation of the Reduced Master Equation in the Mixed Wigner Representation}.\textemdash
Since the dynamics of $\Phi_{u}^{\suW}$, at first-order, depend on the
interaction with the molecule itself,  
Eq.~(\ref{equ:PWIRdcdMstrEqtn}) is not closed from a mathematical viewpoint. 
This can be overcome after introducing a projection operator technique in
the quantum-classical Hilbert--phase-space of the entire molecular-light
field system (see section~S1 in the Supporting Information for an equivalent
treatment in the standard full Hilbert space description). 
For simplicity, consider the second-order regime in the molecule-light
interaction through the Liouvillian superoperators in the interaction
picture of the partial Wigner transform $\hat{\rho}^{\suPW(\suI)}(t) 
= \eu^{\iu\mathcal{L}_{0}^{\suPW} t} \hat{\rho}^{\suPW}(t)$, 
with $\mathcal{L}_{0}^{\suPW} = \mathcal{L}_{\tnS}  +
\mathcal{L}_{\tnR}^{\suW}$. 
Therefore, the quantum-classical Liouville equation in the interaction
picture reads 
\begin{equation}
\label{eq:QC_liouville_equation_interaction}
\frac{\partial}{\partial t}  \hat{\rho}^{\suPW(\suI)}(t)
= -\iu\mathcal{L}_{\tnSR}^{\suPW(\suI)}(t)\hat{\rho}^{\suPW(\suI)}(t),
\end{equation} 
where $\mathcal{L}_{\tnSR}^{\suPW(\suI)}(t) =
\eu^{\iu\mathcal{L}_{0}^{\suPW} t}\mathcal{L}_{\tnSR}^{\suPW}$. 
The factorized interaction Hamiltonian in the interaction representation
assumes the time-dependent form $\hat{H}_{\tnSR}^{\suPW(\suI)}(t) 
= \sum_{u} \hat{K}_u^{(\suI)}(t)\Phi_{u\suW}^{(\suI)}(t)$. 
Thus, the interaction Liouvillian in the interaction representation reads 
$
\mathcal{L}_{\tnSR}^{\suPW(\suI)}(t)\bullet = 
\frac{1}{\hbar}\big[ \hat{H}_{\tnSR}^{\suPW(\suI)}(t), \bullet \big] 
\nonumber
+ \frac{\iu}{2} \big( 
\big\{\hat{H}_{\tnSR}^{\suPW(\suI)}(t), \bullet \big\}_{\tnP} - 
\big\{  \bullet ,\hat{H}_{\tnSR}^{\suPW(\suI)}(t)\big\}_{\tnP} \big) .
$ 
Using the concept of partial trace over the phase-space 
light-field-degrees-of-freedom, and since 
$\tr_{\tnR}\rho_{\tnR}^{\suW(\suI)}(t_0)  = \int \di \mathbf{q} \di 
\mathbf{p}\rho_{\tnR}^{\suW(\suI)}(t_0) = 1$, 
it is then possible to define the quantum-classical projection superoperator
$\mathcal{P}\hat{\rho}^{\suPW(\suI)}(t) = 
\rho_{\tnR}^{\suPW(\suI)}(t_0)\tr_{\tnR}\hat{\rho}^{\suPW(\suI)}(t) $, 
with orthogonal complement $\mathcal{Q} = \hat{1} - \mathcal{P}$. 
The equations of motion for both subspaces read 
\begin{equation}
\label{eq:RdcdMstrEqtnP}
\tr_{\tnR}
\left(\mathcal{P} \frac{\di}{\di t}\hat{\rho}^{\suPW(\suI)}(t)
\right) = 
-\iu\tr_{\tnR} 
\Big(
\mathcal{P}\mathcal{L}_{\tnSR}^{\suPW(\suI)}(t)\rho_{\tnR}^{\suW(\suI)}
(t_0)
\times\hat{\rho}_{\tnS}^{(\suI)}(t) 
+ \mathcal{P} \mathcal{L}_{\tnSR}^{\suPW(\suI)}(t)
\mathcal{Q}\hat{\rho}^{\suPW(\suI)}
(t)
\Big),
\end{equation}
\begin{equation}
\label{eq:RdcdMstrEqtnQ}
\mathcal{Q} \frac{\di \hat{\rho}^{\suPW(\suI)}(t)}{\di t} 
= 
-\iu\Big(
\mathcal{Q} \mathcal{L}_{\tnSR}^{\suPW(\suI)}(t)
\rho_{\tnR}^{\suW(\suI)}(t_0)
\hat{\rho}_{\tnS}^{(\suI)}(t)
+ \mathcal{Q} \mathcal{L}_{\tnSR}^{\suPW(\suI)}(t)
\mathcal{Q}\hat{\rho}^{\suPW(\suI)}(t)
\Big).
\end{equation} 
To second-order, the term 
$\mathcal{Q} \mathcal{L}_{\tnSR}^{\suPW(\suI)}(t)
\mathcal{Q}\hat{\rho}^{\suPW(\suI)}(t)$ 
is disregarded. 
The mixed-Wigner-representation, 
Eqs.~(\ref{eq:RdcdMstrEqtnP}) and \ref{eq:RdcdMstrEqtnQ})  
are analog to Eqs.~(S1) and (S2) in the Supporting Information, respectively.

Considering the factorized initial condition 
$\mathcal{Q}\hat{\rho}^{\suPW(\suI)}(t_0)=0$, the solution of
Eq.~(\ref{eq:RdcdMstrEqtnQ}) reads 
$
\mathcal{Q}\hat{\rho}^{\suPW(\suI)}(t) = -\iu\int_{t_0}^{t}\di t'
\mathcal{Q} \mathcal{L}_{\tnSR}^{\suPW(\suI)}( t') 
\rho_{\tnR}^{\suW(\suI)}(t_0)\hat{\rho}_{\tnS}^{(\suI)}
( t').
$ 
After replacing this result in Eq.~(\ref{eq:RdcdMstrEqtnP}), the first and
second-order reduced master equations for the molecular system read 
\begin{equation}
\label{eq:1st_order_QCME}
\frac{\di}{\di t}\hat{\rho}_{\tnS}^{(\suI)(1)}(t) = 
- \frac{\iu}{\hbar}\sum_u 
\big\langle\Phi_{u}^{\suW(\suI)}(t)\big\rangle\!\left[\hat{K}_u^{(\suI)}(t),
\hat{\rho}_{\tnS}^{(\suI)}(t)\right] ,
\end{equation}
\begin{align}
\label{eq:2nd_order_QCME}
\begin{split}
\frac{\di}{\di t}\hat{\rho}_{\tnS}^{(\suI)(2)}(t) =&  
-\frac{\iu}{\hbar}\sum_u  \big\langle\Phi_{u}^{\suW(\suI)}(t)\big\rangle\!
\left[\hat{K}_u^{(\suI)}(t), 
\hat{\rho}_{\tnS}^{(\suI)}(t)\right]
\\&
- \frac{1}{\hbar^2}\!\int_{t_0}^{t}\!\di t'\sum_{u,v} 
\Big(\mathcal{C}^{\suW}_{u,v}(t, t')\!-\! 
\big\langle\Phi_{u}^{\suW(\suI)}(t)
\big\rangle\big\langle\Phi_{v}^{\suW(\suI)}( t')\big\rangle \Big)\!
\left[\!\hat{K}_u^{(\suI)}(t),\!
\left[\hat{K}_v^{(\suI)}( t'), 
\hat{\rho}_{\tnS}^{(\suI)}( t')\!\right] \right] 
\\&
- \frac{\iu}{2\hbar}\!\int_{t_0}^{t}\!\di t'
\sum_{u,v}\chi_{u,v}^{\suW}(t, t') 
\!\left[\!\hat{K}_u^{(\suI)}(t),\!\left[\hat{K}_v^{(\suI)}( t'), 
\hat{\rho}_{\tnS}^{(\suI)}( t')\!\right]_{\boldsymbol{+}}\!\right]\!,
\end{split}
\end{align}	
where 
\begin{gather}
\mathcal{C}^{\suW}_{u,v}(t, t') \!=\! \frac{1}{2}
	\Big(\!
	\big\langle\!\Phi_{u}^{\suW(\suI)}(t)
	\Phi_{v}^{\suW(\suI)}( t')\!\big\rangle
	\!+\!\big\langle\!\Phi_{v}^{\suW(\suI)}( t') 
	\Phi_{u}^{\suW(\suI)}(t)\!\big\rangle
	\Big),
\label{eq:symmetric_correlation_function}
\\
\chi_{u,v}^{\suW}(t, t') = 
	\expv{\left\{\Phi_{u}^{\suW(\suI)}(t),\Phi_{v}^{\suW(\suI)}( t') 
	\right\}_{\tnP}} ,
\label{eq:linear_response_function}
\end{gather} 
represent the phase-space version of real (symmetrized) part of the
two-time correlation function $g(t,t')$ and the linear response function
\cite{Ing02,CDG98} of the light field operator that couples to the molecular
system. 
$\big\langle\Phi_{u}^{\suW(\suI)}(t)\big\rangle=\tr_{\tnR}
\big[\Phi_{u}^{\suW(\suI)}(t)\rho_{\tnR}^{\suW(\suI)}(t_0)\big]$ represents
the expected value of the light field observable that couples to the
molecular system. 

The reduced master equation in Eq.~(\ref{eq:2nd_order_QCME}) admits 
analyzing the influence of classical or quantum descriptions of light fields 
on the same foot-stage. 
The correlations $\mathcal{C}^{\suW}_{u,v}(t, t')$ and 
$\chi_{u,v}^{\suW}(t, t')$, derived in the quantum-classical projection
operator approach, correspond to the classical limit of the quantities
$\mathcal{C}_{u,v}(t, t')$ and $\chi_{u,v}(t, t')$ in the
Eq.~(S5) of the Supporting Information, respectively. 
The analogy can be understood in terms of the light field two-time
correlation function given by 
\begin{equation}
g^{\suW}_{u,v}(t,t')
= 
\big\langle\Phi_{u}^{\suW(\suI)}(t)\Phi_{v}^{\suW(\suI)}( t')\big\rangle
= \!\int\!\di\mathrm{q}\di\mathrm{p} \,\Phi_{u}^{\suW}(\mathrm{q},\mathrm{p};t)
\Phi_{v}^{\suW}(\mathrm{q},\mathrm{p};t')
\rho_{\tnR}^{\suW(\suI)}(\mathrm{q},\mathrm{p};t_0).
\end{equation} 
In the classical limit, the symmetrized correlation function coincides with
the two-time correlation function 
$\mathcal{C}^{\suW}_{u,v}(t, t') = g^{\suW}_{u,v}(t,t')$. 
The antisymmetrized correlation function is defined through the phase-space Poisson 
bracket $\mathcal{A}^{\suW}_{u,v}(t,t')
=\frac{1}{2\iu}\big\{\Phi_{u}^{\suW(\suI)}(t),\Phi_{v}^{\suW(\suI)}( t')\big\}_{\tnP}$. 
It is useful to express the symmetrized correlation function in the
form given in Eq.~(\ref{eq:symmetric_correlation_function}) for the
comparison with the quantum case discussed below. 
The quantum limit of the Poisson bracket in the phase-space linear response
function corresponds to
$\big\{\Phi_{u}^{\suW(\suI)}(t),\Phi_{v}^{\suW(\suI)}( t')\big\}_{\tnP}
\to \textstyle{\frac{1}{\iu\hbar}}
\big[\hat{\Phi}_{u}^{(\suI)}(t),\hat{\Phi}_{v}^{(\suI)}( t')\big] $.

Since the state of the photon field is aimed not to be measured, up to
second-order in the light-molecule interaction, it is then not possible to
distinguish between the dynamics induced by a classical electric field
having the same statistics as the quantum electric field 
Eq.~(\ref{eq:2nd_order_QCME}) and the actual dynamics generated by the
quantum photon field (see Eq.~(S5) of the Supporting Information). 
However, since the light degrees of freedom evolve independently of the
molecular dynamics, Eq.~(\ref{eq:2nd_order_QCME}) and
Eq.~(S5) of the Supporting Information do not consider the back-action of 
the molecular dynamics on the light field \cite{CV&19}.

This formalism can be extended to (i) generalize the effect of classical
vibrational baths composed of harmonic oscillators or to (ii) include higher
orders in the light-matter interaction; thus, a quantum-classical master
equation of the type Nakajima-Zwanzig or time-convolution-less could be
deduced through this approach, allowing to treat strong coupling 
effects that can lead to deviations from canonical statistics 
\cite{PB13,PYB13,PB14,PT&19,TM&22,ANS23}, 
as recently reported in the modification of ground-state chemical reactivity 
in infrared cavities\cite{AT&23}. 
In addition, if interest is in the dynamics of the light state instead of
the molecule state, the role of the system could be inverted above (see 
Section~S2 of the Supporting Information). 
However, these extensions are beyond the scope of this Letter.

\textit{Traditional Semiclassical Approach}.\textemdash
To fully appreciate the difference between the present approach to the standard 
semiclassical approach and the conundrum posed by the results 
in the \textnormal{NaI} molecule\cite{KBM16,CH&17}, the standard
semiclassical approach is reviewed next. 
The semiclassical Hamiltonian reads 
\begin{equation}
\label{eq:Smclsscl_Hmltnn}
\hat{H}_\mathrm{sc} = 
\hat{H}_{\tnS} - \hat{\boldsymbol{\mu}}\cdot \mathbf{E}(t) 
\end{equation} 
with $\hat{\boldsymbol{\mu}}$ representing the total molecular dipole
operator and the classical electric given by 
$\mathbf{E}(t) = \mathbf{E}_0 \cos(\omega_{\mathrm{c}}t + \phi)$. 
The classical field amplitude $\mathbf{E}_0$ comprises the polarization
state which it is assumed parallel to the total molecular dipole operator. 
The molecular dynamics follow the unitary evolution given by the master
equation 
\begin{align}
\label{eq:semiclassical_ME}
\frac{\mathrm{d}}{\mathrm{d} t}\hat{\rho}^{\mathrm{(\suI)}}_{\mathrm{m}}(t)&=
-\frac{\mathrm{i}}{\hbar}\mathbf{E}(t) \cdot
\left[
-\hat{\boldsymbol{\mu}}^{\mathrm{(\suI)}}(t),\hat{\rho}^{\mathrm{(\suI)}}(t)
\right].
\end{align} 
whose formal solution is given by 
\begin{align}
\label{eq:semiclassical_ME_FS}
\hat{\rho}^{\mathrm{(\suI)}}_{\mathrm{m}}(t)&= 
\hat{\rho}^{\mathrm{(\suI)}}_{\mathrm{m}}(0)
-\frac{\mathrm{i}}{\hbar} \int_0^t \mathrm{d} t' \mathbf{E}(t') \cdot
\left[
-\hat{\boldsymbol{\mu}}^{\mathrm{(\suI)}}(t'),\hat{\rho}^{\mathrm{(\suI)}}(t')
\right].
\end{align} 
For further analysis, It is convenient to replace Eq.~(\ref{eq:semiclassical_ME_FS}) in
Eq.~(\ref{eq:semiclassical_ME}) to have an alternative equation of motion
for the molecule density matrix 
\begin{align}
\label{eq:semiclassical_ME_ME}
\frac{\mathrm{d}}{\mathrm{d} t}
\hat{\rho}^{\mathrm{(\suI)}}_{\mathrm{m}}(t)&=
-\frac{\mathrm{i}}{\hbar}
\left[
- \mathbf{E}(t) \cdot \hat{\boldsymbol{\mu}}^{\mathrm{(\suI)}}(t),
\hat{\rho}^{\mathrm{(\suI)}}_{\mathrm{m}}(0)
\right]
- \frac{1}{\hbar^2} 
\left[ 
-\mathbf{E}(t) \cdot \hat{\boldsymbol{\mu}}^{\mathrm{(\suI)}}(t),
\int_0^t \mathrm{d} t' 
\left[
-\mathbf{E}(t') \cdot \hat{\boldsymbol{\mu}}^{\mathrm{(\suI)}}(t'),
\hat{\rho}^{\mathrm{(\suI)}}_{\mathrm{m}}(t')
\right]
\right].
\end{align} 
To incorporate the classical statistical properties of light (cf. Chap.~4 
in Ref.~\citenum{MW95}), an ensemble average is performed
\begin{align}
\label{eq:semiclassical_ME_ME1}
\frac{\mathrm{d}}{\mathrm{d} t} &
\left<\hat{\rho}^{\mathrm{(\suI)}}_{\mathrm{m}}(t)\right>_\mathrm{light}=
\frac{\mathrm{i}}{\hbar} \left< \mathbf{E}(t) \right>_\mathrm{light} 
\left[
\cdot \hat{\boldsymbol{\mu}}^{\mathrm{(\suI)}} (t),
\hat{\rho}^{\mathrm{(\suI)}}_{\mathrm{m}}(0)
\right] \notag
\\ &
- \frac{1}{\hbar^2} 
\int_0^t \mathrm{d} t' 
\left< \mathbf{E}(t) \cdot \mathbf{E}(t') \right>_\mathrm{light}  
\left[ 
\hat{\boldsymbol{\mu}}^{\mathrm{(\suI)}}(t),
\left[
 \hat{\boldsymbol{\mu}}^{\mathrm{(\suI)}}(t'),
 \hat{\rho}^{\mathrm{(\suI)}}_{\mathrm{m}}(t')
\right]
\right].
\end{align} 
If $\left< \mathbf{E}(t) \right>_\mathrm{light} = 0$, 
as in the case of incoherent light or monochromatic plane wave, then 
\begin{align}
\label{eq:semiclassical_ME_ME2}
\frac{\mathrm{d}}{\mathrm{d} t} 
\left<\hat{\rho}^{\mathrm{(\suI)}}_{\mathrm{m}}(t)\right>_\mathrm{light}&=
- \frac{1}{\hbar^2} 
\int_0^t \mathrm{d} t' \mathcal{C}^{\mathrm{cl}}(t, t')
\left[ 
\hat{\boldsymbol{\mu}}^{\mathrm{(\suI)}}(t),
\left[
\hat{\boldsymbol{\mu}}^{\mathrm{(\suI)}}(t'), 
\hat{\rho}^{\mathrm{(\suI)}}_{\mathrm{m}}(t')
\right]
\right],
\end{align}
with $\mathcal{C}^{\mathrm{cl}}(t, t') = 
\left< \mathbf{E}(t) \cdot \mathbf{E}(t') \right>_\mathrm{light}$. 
For the sake of comparison, consider only one coupling operator in
Eq.~(\ref{eq:2nd_order_QCME}), identify 
$\langle\Phi_{u}^{\suW(\suI)}(t)\big\rangle = \mathbf{E}(t)$ and 
$\hat{K}^{(\suI)}(t) = - \hat{\boldsymbol{\mu}}^{\mathrm{(\suI)}}(t)$  
and reduce for the case 
$\langle \hat{\Phi}_{u}^{\mathrm{(\suI)}}\rangle = 
\langle\Phi_{u}^{\suW(\suI)}(t)\big\rangle = 0$, so that, 
\begin{align}
\label{eq:2nd_order_QCME_1m0_sOp}
\begin{split}
\frac{\di}{\di t}&\hat{\rho}_{\tnS}^{(\suI)(2)}(t) = 
- \frac{1}{\hbar^2}\!\int_{0}^{t}\!\di t' \mathcal{C}^{\suW}(t, t')\!
\!
\left[\!\hat{\boldsymbol{\mu}}^{\mathrm{(\suI)}}(t),\!
\left[\hat{\boldsymbol{\mu}}^{\mathrm{(\suI)}}(t'), 
\hat{\rho}_{\tnS}^{(\suI)}( t')\!\right] \right] 
\\ &
- \frac{\iu}{2\hbar}\!\int_{0}^{t}\!\di t'
\chi^{\suW}(t, t') 
\!\left[\!\hat{\boldsymbol{\mu}}^{\mathrm{(\suI)}}(t),
\!\left[\hat{\boldsymbol{\mu}}^{\mathrm{(\suI)}}(t'), 
\hat{\rho}_{\tnS}^{(\suI)}( t')\!\right]_{\boldsymbol{+}}\!\right]\!.
\end{split}
\end{align}	
The standard semiclassical description of the light-matter interaction in 
Eq.~(\ref{eq:semiclassical_ME}) perfectly replicates the molecular 
dynamics ignited by light, provided that $\mathcal{C}^{\suW}(t, t') = 
\mathcal{C}^{\mathrm{cl}}(t, t')$ and $\chi^{\suW}(t, t') =0$. 
This condition is characteristic of light sources without quantum 
correlations. 
Even in scenarios where $\langle \hat{\Phi}_{u}^{\mathrm{(\suI)}}\rangle = 
0$, it is anticipated that classical light sources can, to a certain 
degree, emulate molecular dynamics as long as $\mathcal{C}^{\suW}(t, t') = 
\mathcal{C}^{\mathrm{cl}}(t, t')$. It should be noted that corrections 
arising from purely quantum correlations of light primarily enter via 
$\chi^{\suW}(t, t')$.

From Eq.(\ref{eq:semiclassical_ME_ME}), it emerges that for quantum states 
like the Fock state where 
$\langle \hat{\Phi}{u}^{\mathrm{(\suI)}}\rangle =
\langle\Phi{u}^{\suW(\suI)}(t)\big\rangle = 0$, a simulation of the 
reduced dynamics of the molecule using classical light sources becomes 
feasible if
$\mathcal{C}^{\suW}(t, t') \approx \mathcal{C}^{\mathrm{cl}}(t, t')$. 
The fidelity of this simulation is contingent upon the light’s 
characteristics, the strength of the coupling, and the inherent complexity 
of the molecular structure. 
Notably, the compatibility of the standard semiclassical description in 
Eq.~(\ref{eq:semiclassical_ME}) with the quantum results for the 
\textnormal{NaI} molecule, as documented in Ref.~\citenum{KBM16}, infers 
the potential accuracy of the effective field description for a variety of 
system models and coupling strengths.

\textbf{\textit{Results for Model Systems}}.\textemdash
The subsequent analysis evaluates the influence of light fields on the 
reduced molecular dynamics in Eq.~(\ref{eq:2nd_order_QCME}) across three 
system models: (i) the Rabi model, (ii) the Dicke model, and (iii) a 
vibronic dimer model. 
The introduction of an effective electric field $\mathbf{E}_\mathrm{eff}$ 
facilitates a discussion on the results from 
Eq.~(\ref{eq:semiclassical_ME_ME2}).

\textit{Rabi Model}.\textemdash
The general framework analyzed in this study is exemplified by considering 
a concrete model system, the Rabi model. 
This model involves a two-level system interacting with a cavity and 
includes the counter-rotating terms, as described by the Hamiltonian 
\begin{equation}
\label{eq:Rabi_Hamiltonian}
\hat{H}_{\mathrm{R}} =  {\textstyle \frac{1}{2}}\hbar \omega_0 \hat{\sigma}_z 
+ \hbar  g \hat{\sigma}_x (\hat{a}^{\dag} +\hat{a})
+ \hbar  \omega_\mathrm{c}\hat{a}^\dagger \hat{a} .
\end{equation} 
Following Eq.~(\ref{eq:2nd_order_QCME_1m0_sOp}), 
the classical description of the light field yields the following master 
equation for the two-level system tracing over the light degrees of freedom 
\begin{align}
\label{eq:2nd_order_QCME_1m0_sOp_Rabi}
\begin{split}
\frac{\di}{\di t}&\hat{\rho}_{\tnS}^{(\suI)(2)}(t) = 
- \frac{1}{\hbar^2}\!\int_{t_0}^{t}\!\di t' \mathcal{C}^{\suW}(t, t')\!
\!
\left[\!\hat{\sigma}_x^{(\suI)}(t),\!
\left[\hat{\sigma}_x^{(\suI)}( t'), 
\hat{\rho}_{\tnS}^{(\suI)}( t')\!\right] \right] 
\\&
- \frac{\iu}{2\hbar}\!\int_{t_0}^{t}\!\di t'
\chi^{\suW}(t, t') 
\!\left[\!\hat{\sigma}_x^{(\suI)}(t),\!\left[\hat{\sigma}_x^{(\suI)}( t'), 
\hat{\rho}_{\tnS}^{(\suI)}( t')\!\right]_{\boldsymbol{+}}\!\right]\!.
\end{split}
\end{align}	
In this context, the quantum equivalent of the phase-space light field 
observable $\Phi^{\suW}(t)$ is defined as 
$\hat{\Phi}^{(\suI)}(t)=\hbar g (\hat{a}^{\dag}\eu^{\iu\omega_\mathrm{c} t} 
+ {\hat{a}\, \eu^{-\iu\omega_\mathrm{c} t})}$, with the 
positive(negative)-frequency component expressed by 
$\hat{\Phi}^{(+)}(t)=\hbar g\,\hat{a}\, \eu^{-\iu\omega_\mathrm{c}t} ,
(\hat{\Phi}^{(-)}(t)=\hbar g\,\hat{a}^{\dag}\eu^{\iu\omega_\mathrm{c} t})$. 
This analysis excludes many-mode-cavity-effects, concentrating solely on one frequency mode $\omega_{\mathrm{c}}$. 
Also, any space dependence of the light-matter coupling strength $g$ is 
not considered \cite{KB&07}. 
Additionally, all Hamiltonians discussed in this section do not include 
the dipole self-energy term \cite{RW&18,SR&20}.

The choice for the classical radiation field in the semiclassical 
approximation corresponds to an electric field whose two-time 
correlation function reproduces the symmetrized correlation 
function $\mathcal{C}(t, t')$ (see Section~S3 of the Supporting Information)
\begin{equation}
\label{E_class}
E_\mathrm{cl} = 2 \hbar g \sqrt{\langle n_\mathrm{c} \rangle
+{\textstyle \frac{1}{2}}}\cos \omega_\mathrm{c} t,
\end{equation}
where $\langle n_\mathrm{c}=\hat{a}^{\dag}\hat{a} \rangle$ represents the average photon number. In the few-photon regime, the semiclassical 
dynamics driven by the classical electric field of Eq.~\ref{E_class} 
deviate significantly from the exact quantum dynamics. Therefore, this 
choice for the electric field displays satisfactory results only 
in the many-photon regime.

Enhancing the results in the few-photon regime from the semiclassical dynamics necessitates 
the incorporation of the emission 
$
\langle \hat{\Phi}^{(+)}(t)\hat{\Phi}^{(-)}(t')\rangle 
= \hbar^2 g^2  \langle \hat{a}^{\dag}\hat{a} +1 \rangle
\exp(-\iu\omega_c(t-t'))
$ 
and absorption 
$
\langle \hat{\Phi}^{(-)}(t)\hat{\Phi}^{(+)}(t')\rangle 
= \hbar^2 g^2  \langle \hat{a}^{\dag}\hat{a}\rangle
\exp(\iu\omega_c(t-t'))
$ 
processes in the molecular system \cite{CDG98,GK05}. 
The expression for the effective electric field is thus
\begin{equation}
E_\mathrm{eff} 
=
2 \hbar g \sqrt{ \langle n_\mathrm{c} \rangle
+ n_\mathrm{a}} \cos(\omega_\mathrm{c} t),
\end{equation}
where $n_\mathrm{a}=0$ ($n_\mathrm{a}=1$) corresponds to the ground ($\ket{g}$) 
(excited $\ket{e}$) state of the two-level system. 
This effective electric field is contingent upon both the initial 
state of the matter and the state of light. 
The semiclassical Hamiltonian,
\begin{equation}
\label{equ:stndrdsmclsscl}
\hat{H}_\mathrm{eff} = \frac{1}{2}\hbar\omega_0 \sigma_z 
+
\hat{\sigma}_x E_\mathrm{eff}, 
\end{equation}
is adept at mimicking the quantum dynamics up to the second order. 
A comparative analysis is conducted with this approach, the quantum-classical 
master-equation formalism 
[Eq.~(\ref{eq:2nd_order_QCME_1m0_sOp_Rabi})], and the full quantum results.

\begin{figure}
\includegraphics[width=\linewidth]{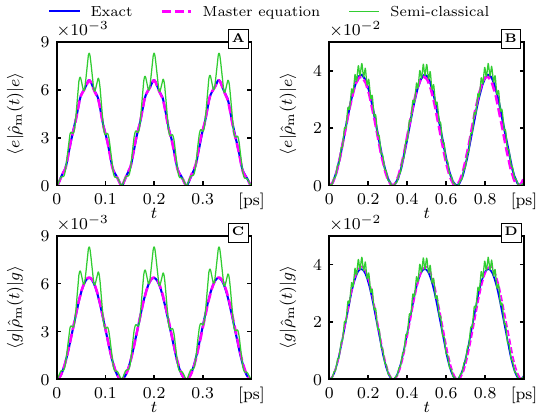}
\caption{\label{fig:RabiModel} 
Populations of the excited (ground) state of the Rabi model as functions  
of time, assuming a Fock state with $n_\mathrm{c}$ photons in the cavity. 
Initially, the entire population is in the ground (excited) state
$\bok{g}{\hat{\rho}_\mathrm{m}(t=0)}{g}=1$ 
($\bok{e}{\hat{\rho}_\mathrm{m}(t=0)}{e}=1)$. 
(A) $\omega_{\mathrm{c}}=0.75\,\omega_0$, $g=0.01\omega_0$ and
$n_{\mathrm{c}}=1$. 
(B) $\omega_{\mathrm{c}}=0.9\,\omega_0$, $g=0.01\omega_0$ and
$n_{\mathrm{c}}=1$. 
(C) $\omega_{\mathrm{c}}=0.75\,\omega_0$, $g=0.01\omega_0$ and
$n_{\mathrm{c}}=0$. 
(D) $\omega_{\mathrm{c}}=0.9\,\omega_0$, $g=0.01\omega_0$ and
$n_{\mathrm{c}}=0$. 
Color coding is shown on top. 
}
\end{figure}

\begin{figure}
\includegraphics[width=\linewidth]{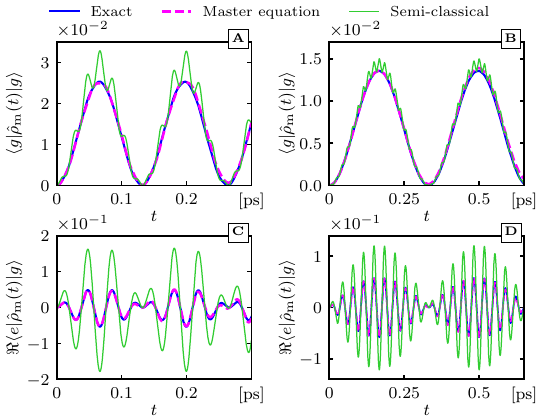}
\caption{\label{fig:RabiModel2} 
Ground state populations and ground-excited state coherences of the Rabi 
model as functions of time, considering different Fock state 
superpositions: (A) and (C) $\omega_{\mathrm{c}}=0.75\omega_0$, 
$g=0.015\omega_0$ and 
$\ket{\psi_{\mathrm{c}}} = \sqrt{0.2}\ket{0}+\sqrt{0.8}\ket{1}$. 
(B) and (D)  $\omega_{\mathrm{c}}=0.9\,\omega_0$, $g=0.0025\omega_0$ and 
$\ket{\psi_{\mathrm{c}}} = \sqrt{0.5}\ket{4}+\sqrt{0.5}\ket{5}$. 
Initially, the entire population is in the excited state 
$\bok{e}{\hat{\rho}_\mathrm{m}(t=0)}{e}=1$. 
Color coding is shown on top.
}
\end{figure}

The forthcoming simulations assume that the phase-space 
correlations\cite{Car99,Car08} in 
Eq.~(\ref{eq:2nd_order_QCME_1m0_sOp_Rabi}) are aligned with their 
quantum-mechanical counterparts. 
Figures~\ref{fig:RabiModel} and ~\ref{fig:RabiModel2} illustrate the 
dynamics of the ground and excited electronic state populations of the two-level system. 
The exact quantum (blue), quantum-classical master equation (magenta), and 
standard semiclassical (green) dynamics, induced by the corresponding 
effective electric field, are compared. 
The analysis includes various states of the light field exhibiting 
non-classical signatures\cite{MW95,GK05,Aga12}. 
The numerical integration of Eq.~(\ref{eq:2nd_order_QCME_1m0_sOp_Rabi}) 
extends beyond the secular approximation\cite{DT&18,DDK19} and encompasses 
non-Markovian effects due to the non-locality in time of the 
dynamics\cite{PYB13}. 

Initially, the case of a Fock state 
$\ket{\psi_\mathrm{c}} = \ket{n_\mathrm{c}}$, with $n_\mathrm{c}$ photons 
in the cavity, is examined; the initial condition is 
$\hat{\rho}_{\tnR}(t_0) = \proj{n_\mathrm{c}}{n_\mathrm{c}}$. 
The Wigner representation of this state features negative regions, 
indicative of a non-classical character\cite{MW95,GK05,Aga12}. 
For this state of the light field, the expected value of the light field 
operator is $\big\langle\hat{\Phi}^{(\suI)}(t)\big\rangle=0$. 
The symmetrized correlation function is expressed as 
$\mathcal{C}(t, t') = \hbar^2 g^2 (2 n_\mathrm{c} + 1) 
\cos(\omega_\mathrm{c} (t -  t'))$, and the antisymmetrized correlation 
function is 
$\mathcal{A}(t, t') = - \hbar^2 g^2 \sin(\omega_\mathrm{c} (t -  t'))$. 
The value of $\mathcal{A}(t, t')$ is independent of the cavity state and 
is determined by the commutation relation 
$[\hat{a},\hat{a}^\dagger]=\hat{1}$ (refer to Section~S3 of the Supporting 
Information).

Figures~\ref{fig:RabiModel}(A) and \ref{fig:RabiModel}(B) illustrate the 
population dynamics of the excited electronic state $\bok{e}
{\hat{\rho}_\mathrm{m}(t)}{e}$ for a one-photon Fock state 
$\ket{\psi_\mathrm{c}}=\ket{1}$ in the cavity. 
Two distinct frequencies, $\omega_{\mathrm{c}}=0.75\,\omega_0$ 
(Fig.~\ref{fig:RabiModel}(A)) and $\omega_{\mathrm{c}}=0.9\omega_0$ 
(Fig.~\ref{fig:RabiModel2}(B)), and a coupling strength of 
$g=0.01\omega_0$ are considered. 
The assumption here is an initial total population in the ground 
electronic state, given by $\bok{g}{\hat{\rho}_\mathrm{m}(t=0)}{g}=1$. 
Conversely, Figures~\ref{fig:RabiModel}(C) and \ref{fig:RabiModel}(D) 
depict the population dynamics of the ground electronic state $\bok{g}
{\hat{\rho}_\mathrm{m}(t)}{g}$ for a zero-photon Fock state 
$\ket{\psi_\mathrm{c}}=\ket{0}$ (vacuum state) in the cavity. 
The same frequencies and coupling strength as the prior case are 
considered, and the initial total population is assumed to be in the 
excited electronic state, expressed as $\bok{e}{\hat{\rho}_\mathrm{m}(t=0)}
{e}=1$. 

A comparison with the exact quantum results reveals that the dynamics 
described by the quantum-classical master equation are more accurate than 
those obtained from the standard semiclassical approach. 
The latter exhibits an oscillatory behavior that is overestimated compared 
to the exact quantum results, a discrepancy attributable to the exclusion 
of the complex part of the two-time correlation function in the effective 
electric field. 
Consider now a Fock state superposition 
$\ket{\psi_{\mathrm{c}}} = c_n\ket{n}+c_{n+1}\ket{n+1}$. 
The symmetrized correlation function for this state is given by 
$
\mathcal{C}(t, t') = \hbar^2 g^2 \big(|c_n|^2(2 n_\mathrm{c} + 1) + 
|c_{n+1}|^2(2 n_\mathrm{c} + 3) \big) \cos \omega_\mathrm{c} (t -  t'),
$ 
while the antisymmetrized correlation function $\mathcal{A}(t, t')$ 
remains constant at $- \hbar^2 g^2 \sin \omega_\mathrm{c}(t -  t')$ as it 
is independent of the light field state. This state also yields a non-zero 
expected value of the light field operator 
$
\big\langle\hat{\Phi}^{(\suI)}(t)\big\rangle = 2\hbar 
g\sqrt{n+1}c_nc_{n+1}\cos\omega t.
$
This light field state exhibits non-trivial quantum characteristics, as 
indicated by a negative value of the Mandel parameter\cite{MW95,GK05,Aga12}.

Fig.~2 illustrates the ground state population (Cases A and B) 
and ground-excited state coherence (Cases C and D) dynamics of the 
two-level system. 
In Figures 2(A) and 2(C), a superposition of the vacuum and 
one-photon Fock states 
$\ket{\psi_{\mathrm{c}}} = \sqrt{0.2}\ket{0}+\sqrt{0.8}\ket{1}$ is 
considered, with a frequency $\omega_{\mathrm{c}}=0.75\omega_0$ and a 
coupling strength $g=0.015\omega_0$. 
In contrast, Figures 2(B) and 2(D) involve a light field state 
$\ket{\psi_{\mathrm{c}}} = \sqrt{0.5}\ket{4}+\sqrt{0.5}\ket{5}$, frequency 
$\omega_{\mathrm{c}}=0.9\omega_0$, and a coupling strength 
$g=0.0025\omega_0$. 
One might expect that the non-classicality of the Fock state's Wigner 
representation, indicated by negative regions in the phase space, would 
result in dynamics distinct from those produced by classical sources. 
However, this expectation does not hold, corroborated by previous works 
\cite{KBM16,CH&17,CV&19}. 
A comparison to the exact quantum results reveals that the 
quantum-classical master equation offers more accurate dynamics than the 
standard semiclassical approach, which exhibits an overestimated 
oscillatory behavior.

\begin{figure}
\includegraphics[width=\linewidth]{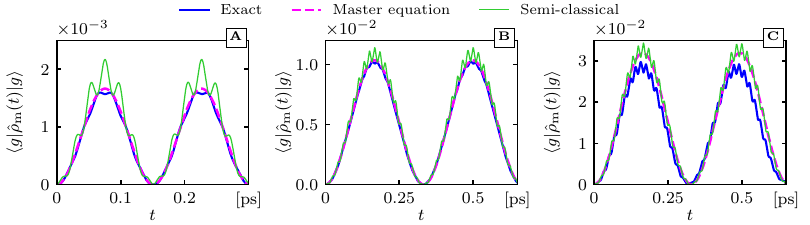}
\caption{\label{fig:RabiModel3} 
Ground state populations of the Rabi model as functions of time, considering 
different squeezed vacuum states: 
(A) $\omega_{\mathrm{c}}=0.75\omega_0$, $g=0.005\omega_0$ and 
$\ket{\psi_{\mathrm{c}}} = \ket{\xi;r=0.2}$. 
(B) $\omega_{\mathrm{c}}=0.9\omega_0$, $g=0.005\omega_0$ and 
$\ket{\psi_{\mathrm{c}}} = \ket{\xi;r=0.2}$. 
(C) $\omega_{\mathrm{c}}=0.9\omega_0$, $g=0.005\omega_0$ and 
$\ket{\psi_{\mathrm{c}}} = \ket{\xi;r=1.2}$. 
Initially, the entire population is in the excited state 
$\bok{e}{\hat{\rho}_\mathrm{m}(t=0)}{e}=1$.  
Color coding is shown on top.
}
\end{figure}

Considering a squeezed vacuum state 
$\ket{\psi_{\mathrm{c}}} = \ket{\xi}$ \cite{BP02,Aga12}, where $\varphi=0$ 
and thus $\xi=r$, and $r$ denotes the squeezing parameter (see section S3 
of the Supporting Information), the symmetrized correlation function 
becomes
$
\mathcal{C}(t, t') = \hbar^2 g^2 \big((2\sinh^2 r + 1) \cos 
\omega_\mathrm{c} (t - t') - 2(\cosh r \sinh r) 
\cos \omega_\mathrm{c} (t + t')\big),
$
with the expected value of the light field operator given by 
$\big\langle\hat{\Phi}^{(\suI)}(t)\big\rangle=0$. 
The non-classical nature of the squeezed vacuum state is highlighted by a 
negative squeezing parameter 
$ S = -\frac{1}{2} (1 - \mathrm{e}^{-2r})$ \cite{Aga12}.

Figures 3(A) and 3(B) illustrate the 
ground state population of the two-level system for a squeeze parameter 
$r=0.2$. 
The quantum-classical master equation offers superior results compared to 
the standard semiclassical approach. 
The latter exhibits an exaggerated oscillatory behavior in contrast with 
the exact quantum results. 
However, it's noteworthy that for higher values of $r$, even under weak 
light-matter coupling, both the quantum-classical master equation and the 
standard semiclassical approaches fail to accurately model the 
molecular-only dynamics (see Figure 3C). 
This deviation underscores a distinct quantum signature within the 
molecular-only dynamics, diverging significantly from the dynamics 
influenced by classical light sources \cite{TPS18}.

\begin{figure}
\includegraphics[width=\linewidth]{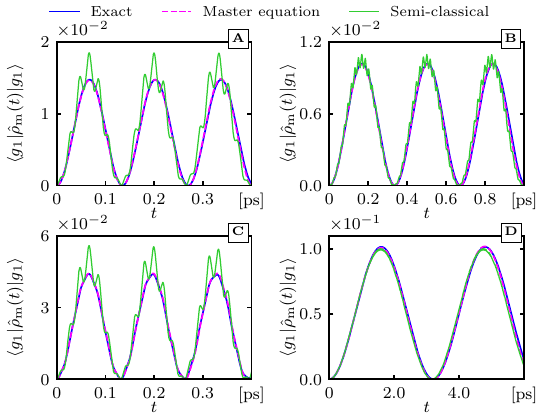}
\caption{\label{fig:Dicke} 
Ground state populations of the site one as functions of time in the Dicke
model of 4 atoms initially in their excited states 
It is assumed a Fock state with $n_{\mathrm{c}}$ photons in the cavity. 
(A) $\omega_{\mathrm{c}}=0.75\,\omega_0$, $g=0.015\omega_0$ and 
$n_{\mathrm{c}}=0$. 
(B) $\omega_{\mathrm{c}}=0.9\,\omega_0$, $g=0.005\omega_0$ and 
$n_{\mathrm{c}}=0$. 
(C) $\omega_{\mathrm{c}}=0.75\,\omega_0$, $g=0.008\omega_0$ and 
$n_{\mathrm{c}}=10$. 
(D) $\omega_{\mathrm{c}}=0.99\,\omega_0$, $g=0.0005\omega_0$ and 
$n_{\mathrm{c}}=10$. 
Color coding is shown on top.
}
\end{figure}

\textit{Dicke Model}.\textemdash
The Dicke model, describing an ensemble of \(N\) two-level atoms 
collectively coupled to a single quantized cavity mode, is renowned for 
its ground-state transition to a superradiant phase \cite{KR&19, DDK19}. 
The Hamiltonian for the Dicke model is given by
\begin{equation}
\label{eq:Dicke_Hamiltonian}
\hat{H}_{\mathrm{D}} = {\textstyle \frac{1}{2}}\hbar\omega_0 
\sum_{i=1}^{N}\hat{\sigma}_{z(i)} 
+ \hbar \omega_\mathrm{c}\hat{a}^\dagger\hat{a} 
+ \hbar g \sum_{i=1}^{N}\hat{\sigma}_{x(i)} (\hat{a}^{\dag} + \hat{a}).
\end{equation}
This model, a cornerstone of many-body dynamics in quantum optics, has 
recently been explored within the realm of atom-only dynamics utilizing 
the non-secular Redfield master equation approach \cite{DDK19}.

Fig.~\ref{fig:Dicke} illustrates the ground state population of one of the 
four two-level atoms, all initially in their excited states and 
considering a Fock state in the cavity.  
Results for a quantized cavity mode corresponding to a vacuum state 
\(\ket{\psi_\mathrm{c}} = \ket{0}\) with frequencies 
\(\omega_{\mathrm{c}}=0.75\,\omega_0\) and 
\(\omega_{\mathrm{c}}=0.9\,\omega_0\), and coupling strengths 
\(g=0.015\omega_0\) and \(g=0.005\omega_0\) are presented in 
Fig.~\ref{fig:Dicke}(A) and \ref{fig:Dicke}(B) respectively. 
Fig.~\ref{fig:Dicke}(C) depicts a scenario where the quantized cavity mode 
is in a Fock state of ten photons \(\ket{\psi_\mathrm{c}}=\ket{10}\), with 
a frequency of \(\omega_{\mathrm{c}}=0.75\,\omega_0\), and a coupling 
strength of \(g=0.008\omega_0\). 
Fig.~\ref{fig:Dicke}(D) considers a cavity mode also in a ten-photon Fock 
state but in resonance with the two-level atom excitation energy, 
\(\omega_{\mathrm{c}}=0.99\,\omega_0\), and a significantly reduced 
coupling strength \(g=0.0005\omega_0\).

Across all considered scenarios, the quantum-classical master equation 
approach outperforms the standard semiclassical dynamics in approximation 
accuracy to the exact quantum dynamics.

\textbf{\textit{Conclusions}}.\textemdash
This study has presented a second-order quantum-classical master equation 
formalism, laying down the criteriafor classical light field states to emulate the 
effects of quantum light sources on molecular systems. 
The comparative analysis was anchored in the simulation of molecular-only 
dynamics, juxtaposing the quantum-classical, quantum-exact, and 
semiclassical approaches across various systems within the scope of 
molecular polaritonics.

A discernible alignment is observed with the quantum-classical approach 
resonating more accurately with the quantum exact molecular-only dynamics 
at the second-order in light-matter interaction relative to the 
semiclassical approach. 
A pivotal realization surfaces that the induced dynamics between a 
classical electric field mirroring the statistics of the quantum electric 
field are indistinguishable.

The implication is profound, underscoring that the intrinsic non-classical
character of a light quantum state does not predicate a distinctive 
molecular dynamic relative to excitations emanating from a classical 
source, a phenomenon evident in the behavior of Fock states. 
This observation finds corroboration in other investigative forays into 
polaritonic chemistry.

A curious capability emerges, where classical light fields exterior to the 
cavity convincingly replicate effects intrinsic to the cavity’s confines. 
A caveat, however, surfaces in scenarios marked by weak light-matter 
coupling. 
Here, a vacuum squeezed state, characterized by a high value of the 
squeeze parameter \(r\), precipitates a departure where both the 
quantum-classical master equation and the standard semiclassical approach 
falter in capturing the nuanced molecular-only dynamics. 
This underscores the emergence of complex, non-trivial quantum 
molecular-only dynamics.

An avenue of exploration and expansion beckons, where the 
quantum-classical open quantum system approach can be extrapolated to 
embrace multiple cavity modes. 
This integration promises to weave in the often sidestepped dissipation 
and decoherence effects, injecting a layer of realism often overlooked in 
the prevailing light-matter interaction wave function methodologies.

\begin{acknowledgement}
Discussions with Profs. Paul Brumer, Jos\'e Luis Sanz-Vicario, Alessandro Sergi 
and Johan F. Triana are acknowledged with pleasure. 
L.F.C. acknowledges financial support from the \emph{Departamento
Administrativo de Ciencia, Tecnolog\'ia e Innovaci\'on} --COLCIENCIAS--. 
This work was supported by the \emph{Vicerrector\'ia de Investigaci\'on} at Universidad 
de Antioquia, Colombia.
\end{acknowledgement}

\begin{suppinfo}
\label{suppinfo}
Second Order Master Equation; Light-field dynamics; Light field correlation functions.
\end{suppinfo}

\bibliography{qtccce}

\end{document}